# Effects of oxygen-reducing atmosphere annealing on LaMnO$_3$ epitaxial thin films

W S Choi[1], Z Marton[2], S Y Jang[1], S J Moon[1], B C Jeon[1], J H Shin[1], S S A Seo[2], T W Noh[1], K Myung-Whun[3], H N Lee[2,a)], and Y S Lee[4,b)]

[1]ReCOE & FPRD, Department of Physics and Astronomy, Seoul National University, Seoul 151-747, Korea

[2]Materials Science and Technology Division, Oak Ridge National Laboratory, Oak Ridge, Tennessee 37831, USA

[3]Department of Physics, Chonbuk National University, Jeonju 561-756, Korea

[4]Department of Physics, Soongsil University, Seoul 156-743, Korea

a)Electronic mail: hnlee@ornl.gov
b)Electronic mail: ylee@ssu.ac.kr

**Abstract**

We investigated the effects of annealing on LaMnO$_3$ epitaxial thin films grown by pulsed laser deposition and propose an efficient method of characterizing their stoichiometry. Structural, magnetic, and optical properties coherently indicate non-stoichiometric ferromagnetic and semiconducting phases for as-grown LaMnO$_3$ films. By annealing in an oxygen-reducing atmosphere, we recovered the antiferromagnetic and insulating phases of bulk-like stoichiometric LaMnO$_3$. We show that non-destructive optical spectroscopy at room temperature is one of the most convenient tools for identifying the phases of LaMnO$_3$ films. Our results serve as a prerequisite in studying LaMnO$_3$ based heterostructures grown by pulsed laser deposition.

Perovskite LaMnO$_3$ (LMO) has recently attracted renewed attention as a building block of multilayered heterostructures. In its bulk phase, LMO is an *A*-type antiferromagnetic (AFM) insulator and hole doping by substituting Sr or Ca for La brings it into a ferromagnetic (FM) semiconducting phase [1-4]. Partially due to its proximity to a FM semiconducting phase, LMO is often used as a component in multilayered heterostructures to investigate intriguing electric/magnetic/optical properties. For example, the LMO and SrMnO$_3$ interface shows both an anomalous magnetic ground state and metal insulator transition that have been attributed to an interface proximity effect [5, 6]. An optical magnetoelectric effect has also been observed in a tricolor superlattice composed of LMO, SrMnO$_3$, and LaAlO$_3$ [7]. In addition, LMO/SrTiO$_3$ (STO) superlattices have exhibited a FM signal, although this is not yet clearly understood [8].

To properly address the physics of LMO-based heterostructures, the physical properties of the LMO layer must be primarily understood. It is known that LMO easily adopts excess oxygen from its stoichiometric phase. Since the perovskite phase does not allow interstitial oxygen, it has cation vacancies instead. Indeed, there have been several reports of LMO films exhibiting FM and semiconducting behaviour where cation vacancies have been proposed as the main cause [9-11]. Although such non-stoichiometric LMO films might give misleading information when studying the intrinsic properties of the heterostructures, systematic efforts to retrieve and investigate the bulk-like phase in as-grown epitaxial LMO films have not been sufficiently pursued.

We investigated the effects of annealing on the physical properties of epitaxial LMO thin films deposited by pulsed laser deposition (PLD). As-grown films showed a FM and semiconducting phase, whereas after high temperature annealing in an oxygen-reducing atmosphere, the film recovered its bulk-like AFM and insulating phases. Using optical spectroscopy, a very effective and convenient tool for characterizing the oxygen stoichiometry of oxide films at room temperature, we developed a precise understanding of the difference between the phases before and after annealing.

We fabricated LMO epitaxial thin films by PLD on 001 STO single-crystal substrates. These samples were grown at 700°C in 10 mTorr of oxygen; the layer count as grown was monitored by observing the reflection high-energy electron diffraction specular spot. Some of the as-grown films were then post-annealed in an oxygen-reducing atmosphere, *i.e.*, in a forming gas (4% H + 96% Ar), at 500–900°C for 1–3 hours where it was expected some oxygen would be extracted from the films. We confirmed the physical properties of the STO substrate did not change upon annealing. It should be noted that the film's physical properties changed abruptly when annealed above 700°C for more than one hour. Film thicknesses were about 30 nm, as calculated from the clear thickness fringe reflections of x-ray diffraction (XRD) measurements taken at the synchrotron radiation source (11.2 keV) at the 10C1 beam line of the Pohang Light Source (PLS). These also indicated good crystallinity in the films. Temperature dependent magnetizations (M(*T*)) (zero-field cooled) were measured using a SQUID magnetometer (Quantum Design). For the spectroscopic study of the LMO film electronic structure, we measured near-normal-incident reflectance and transmittance spectra in the wide photon energy range of 0.07–5.9 eV. We used a Fourier transform infrared spectrometer (Bruker IFS66v/S) and a grating-type spectrophotometer (CARY 5G) at 0.07–1.2 and 0.4–5.9 eV, respectively. We calculated the in-plane optical conductivity spectra $\sigma(\omega)$ at 0.3–3.2 eV, where the STO substrate was transparent, using an intensity-transfer-matrix method numerical-iteration process [12].

Figures 1(a) and (b) show XRD $\theta$-$2\theta$ scans for the as-grown and annealed LMO films, respectively, taken to investigate the effect of annealing on the structural properties of LMO films. In Fig. 1(a), the as-grown film shows a weak 002 reflection that looks like a shoulder on the prominent substrate 002 peak. Conversely, the annealed film in Fig. 1(b) shows a pronounced peak with an increased *c*-axis lattice constant. Taking the difference between the peak positions, we estimated that the LMO film *c*-axis lattice constant increased by about 1% with annealing. Reciprocal space maps around the 103 STO peak of the as-grown and annealed LMO film, respectively, were used to characterize the in-plane strain state (data not shown). While the as-grown LMO film peak was at *l* = 3.00, the annealed LMO film peak was at *l* = 2.975, again indicating an increase in the *c*-axis lattice constant. On the other hand, the in-plane *h* value of the peak was fixed at 1.00, confirming that the in-plane lattices for both the as-grown and the annealed LMO films were coherently strained to the substrate. The LMO film unit cell volume expanded from 59.65 Å$^3$ to 60.59 Å$^3$ with annealing. Notably, the annealed LMO film unit cell volume was closer to that of a high-quality single-crystal, which can be as large as 60.95 Å$^3$ [1], This indicates that by removing the oxygen, forming gas annealing causes the LMO film unit cell volume to increase and approach the bulk value.

Figure 2 with the M(*T*) of the as-grown and annealed LMO films, shows the effects of annealing on the magnetic properties of the films. The M(*T*) curve for the as-grown film increased steeply as the temperature decreased below about 200 K, as is typical for an FM ordering. The saturated magnetization reached about 3.5 $\mu_B$/Mn at 10 K. Conversely, for the annealed film, the M(*T*) curve increased slowly with decreasing temperature, and the saturated magnetization was drastically reduced to about 0.5 $\mu_B$/Mn. Moreover, the onset temperature of the magnetization was reduced to about 145 K. While the curvature of the M(*T*) curve signaled a weak FM ordering below the onset temperature, the drastically suppressed magnetic moment was not compatible with the perfect FM ordering of $Mn^{3+}$ ion local spins. Note also that the onset temperature was very close to the AFM ordering temperature (~140 K) of the single-crystal LMO [1].

We also measured the magnetic field-dependent magnetization (M-H) curve at 10 K, as shown in the inset of Fig. 2. The M-H curve also showed drastically reduced magnetization for the annealed film where the ~2.2 $\mu_B$/Mn remnant magnetization decreased to ~0.2 $\mu_B$/Mn upon annealing. While the M-H curve for as-grown film was compatible with the normal FM ground state, the miniscule remnant magnetic moment for the annealed film suggested that the FM ground state of the film was possibly due to canting of the antiferromagnetically coupled spins. Such a reduced FM moment has typically been observed in single-crystals with canted AFM spins [1]. The dramatically reduced remnant magnetic moment and the reduced onset temperature of the magnetization suggested that the bulk-like AFM ground state was revived with annealing in an oxygen reducing condition.

Figure 3 shows $\sigma(\omega)$ of as-grown and annealed LMO films at room temperature. For comparison, data reproduced from the literature for LMO and $La_{0.9}Sr_{0.1}MnO_3$ single-crystals are also shown [3]. The $\sigma(\omega)$ of as-grown and annealed LMO films had similar spectral features to doped (*i.e.*, $La_{0.9}Sr_{0.1}MnO_3$) and undoped LMO single-crystals, respectively. Careful comparison of $\sigma(\omega)$ for the LMO films with single-crystalline phases provided a complementary understanding of the forming gas annealing mechanism that recovered the bulk-like LMO phase.

Similarly to the LMO single-crystal, the $\sigma(\omega)$ of the annealed LMO film showed an $\omega^2$-like rapid increase just above the optical gap with a ripple-like shaped peak centered at ~2.3 eV. The upturn starting at ~3 eV, observed in all of the films and single crystals, indicated the initiation of the charge-transfer transition from the oxygen to the Mn ions. The ~2.3 eV peak has been previously attributed to the photo-excitation of electrons in the $Mn^{3+}$ ion $e_g$ orbitals in stoichiometric LMO [13, 14]. The remarkable resemblance between the annealed film and LMO single crystal indicated that the annealed film had a stoichiometric phase and that the peak at ~2.3 eV in the film was due to the photo-excitation of the $Mn^{3+}$ ion $e_g$ orbitals.

Compared to the annealed film, the $\sigma(\omega)$ of the as-grown LMO film showed three characteristic features: a slightly decreased optical gap, a suppression of the ~2.3 eV peak as well as the ripple-like pattern, and a spectral weight increase at ~1.7 eV. The $\sigma(\omega)$ of the hole-doped manganite $La_{0.9}Sr_{0.1}MnO_3$ single-crystal showed similar spectral features. While the stoichiometric LMO crystal has only $Mn^{3+}$-ions, Sr-doping introduces $Mn^{4+}$-ions into the Mn-sublattice. $Mn^{4+}$-ions accompany holes in the $e_g$ orbitals, which results in semiconducting FM coupling between the local spins via double exchange. Moreover, the holes become polaronic carriers, which cause the development of a mid-gap state below the optical gap in LMO [15]. The corresponding development of the mid-gap state in the as-grown LMO film suggested the presence of $Mn^{4+}$-ions in the film, probably due to cation vacancies induced during the highly energetic PLD process. In the inset of Fig. 3, we show the temperature dependent resistivity ($\rho$(T)) for the as-grown and annealed LMO films. While the annealed LMO film had an extremely high $\rho$ with $\rho$(T) behaviour typical for insulators, the as-grown film showed a lower $\rho$ by more than two orders of magnitude at room temperature. Moreover, we could observe an anomaly at around the onset temperature of the FM ordering, which seems to originate from the double exchange interaction due to the $Mn^{4+}$-ions in the as-grown film.

From the resemblance of $\sigma(\omega)$ in the as-grown LMO film to that of the $La_{0.9}Sr_{0.1}MnO_3$ single-crystal, we estimated the amount of cation vacancies in the film and identified the phase of the as-grown film. In the $La_{0.9}Sr_{0.1}MnO_3$ single crystal, 10% of the $Mn^{3+}$-ions became $Mn^{4+}$-ions to match the chemical valence. According to the relationship $(La^{3+}_{1-x}[\ ]_x)(Mn^{3+}_{1-7x}Mn^{4+}_{6x}[\ ]_x)O_3^{-2}$, the value of $x$ should be 1/60 to induce a 10% transition to $Mn^{4+}$-ions. This suggests that under the assumption that there are simultaneously both La and Mn vacancies, about 1.67% of the unit cells should contain cation vacancies. It has previously been reported that when the value of $x$ is larger than about 0.03, the LMO single-crystal undergoes a phase transition from orthorhombic to rhombohedral, and the spin ordering becomes completely FM [1]. Since $x < 0.03$ in our as-grown LMO film, and the strain from the substrate is likely to suppress any possible structural phase transition, we concluded that the film could have an orthorhombic structure with a reasonable amount of FM components.

The $c$-axis lattice constant increase upon annealing can also be explained by the valence state of the Mn-ions. Suppose that the as-grown LMO has $Mn^{4+}$ ions along with $Mn^{3+}$ ions. Since $Mn^{3+}$ ions (6.5 pm) are larger than $Mn^{4+}$ ions (5.4 pm) [16], the annealed LMO film, consisting primarily of $Mn^{3+}$ ions, should have the larger observed lattice constant.

The XRD, magnetization, and $\sigma(\omega)$ data coherently demonstrate that annealing in an oxygen reducing atmosphere is necessary to reduce the $Mn^{4+}$ ions in as-grown films so as to recover the bulk-like magnetic and optical properties. Optical spectroscopy has a particular advantage as the measurement as it can be done simply and quickly at room temperature. By comparing with spectra in the literature, it could also provide a quantitative evaluation of the rough estimate of the amount of cation vacancies.

In summary, we investigated the effects of forming gas annealing on various physical properties of LMO epitaxial thin films grown by PLD. We found that annealing in an oxygen-reducing atmosphere at more than 700°C brought the as-grown films to a bulk-like stoichiometric phase of LMO. Our study emphasized that the LMO films deposited by PLD are quite vulnerable and that their stoichiometry should be carefully characterized. We also demonstrated that non-destructive optical spectroscopy can be used as an effective and convenient tool to characterize the stoichiometry of LMO films. The optical spectroscopy methodology can be further extended to effectively characterize the stoichiometry of numerous other oxide films, such as $LaTiO_3$ and $LaVO_3$, and might also be expanded to examine other manganite oxides such as $Pr_{(1-x)}Ca_xMnO_3$ [17].

**Acknowledgements**
We thank J.-Y. Kim for valuable discussion. This research was supported by Basic Science Research Program Through the National Research Foundation of Korea (NRF) funded by the Ministry of Education, Science and Technology (No. 2009-0080567). The experiments at PLS were partially supported by POSCO. The work at ORNL was sponsored by the Division of Materials Sciences and Engineering, U. S. Department of Energy. YSL was supported by the Soongsil University Research Fund.

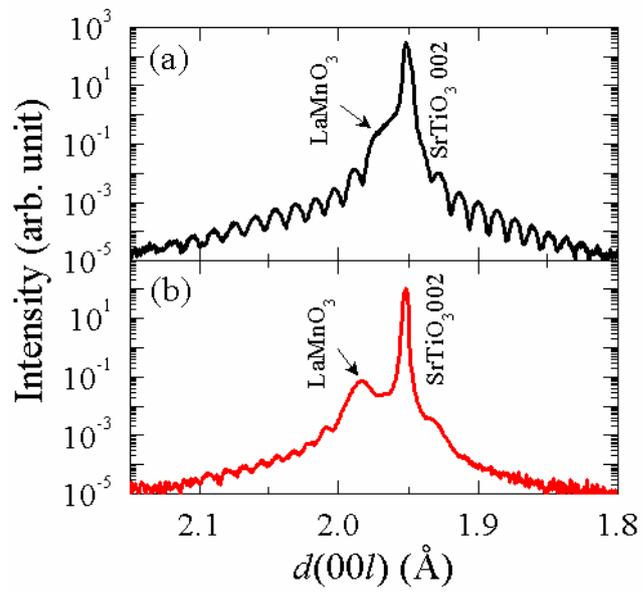

Figure 1. X-ray diffraction $\theta$-$2\theta$ patterns near the 002 Bragg condition of (a) as-grown and (b) forming gas annealed LaMnO$_3$ films at 900°C for 1 hour.
(This figure is in colour only in the electronic version)

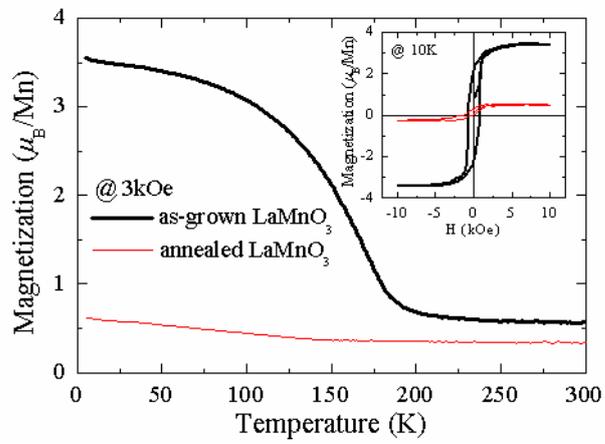

Figure 2. (Color online) Magnetization as a function of temperature (zero-field cooled, while warming) with a 3kOe magnetic field applied along the *a*-axis for as-grown (thick black line) and annealed (thin red line) $LaMnO_3$ films. Inset shows the magnetization as a function of magnetic field at 10 K.
(This figure is in colour only in the electronic version)

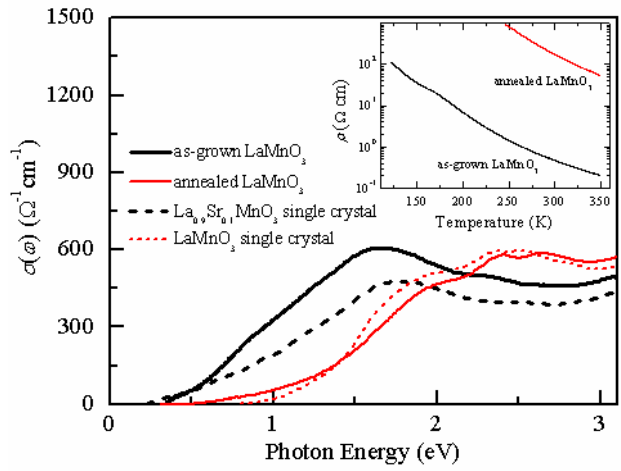

Figure 3. (a) $\sigma(\omega)$ for the as-grown (thick black solid line) and annealed (thin red solid line) LaMnO$_3$ films. For comparison, $\sigma(\omega)$ for La$_{0.9}$Sr$_{0.1}$MnO$_3$ (thick black dotted line) and LaMnO$_3$ (thin red dotted line) single crystals are also shown. [Ref. 3] The inset shows the temperature dependent resistivity for as-grown (thick black line) and annealed (thin red line) LaMnO$_3$ films. (This figure is in colour only in the electronic version)